# Testing quantum transport without decoherence

E. Flores-Olmedo[1], A. M. Martínez-Argüello[2], M. Martínez-Mares[2], G. Báez[1], J. A. Franco-Villafañe[3] and R. A. Méndez-Sánchez[3]

[1] *Departamento de Ciencias Básicas, Universidad Autónoma Metropolitana-Azcapotzalco, Av. San Pablo 180, Col. Reynosa Tamaulipas, 02200 México Distrito Federal, Mexico*

[2] *Departamento de Física, Universidad Autónoma Metropolitana-Iztapalapa, Apartado Postal 55-534, 09340 México Distrito Federal, Mexico*

[3] *Instituto de Ciencias Físicas, Universidad Nacional Autónoma de México, Apartado Postal 48-3, 62210 Cuernavaca Mor., Mexico*

**The coherent electronic transport phenomena through quantum devices is difficult to observe due to thermal smearing and dephasing, the latter induced by inelastic scattering by phonons or impurities. In other wave systems, the temperature and dephasing may not destroy the coherence and can then be used to observe the underlying wave behaviour of the coherent quantum phenomena. Here, we observe coherent transmission of mechanical waves through a two-dimensional elastic Sinai billiard with two waveguides. The flexural-wave transmission performed by non-contact means, shows the transmission quantization when a new mode becomes open. In contrast with the quantum counterpart, these measurements agree with the theoretical predictions of the simplest model at zero temperature highlighting the universal character of the transmission fluctuations.**

The electronic transport through clean conductors, ranging from nanometric to micrometric scale, at low temperature is expected to be coherent and well described by quantum mechanics. In that situation, the electrons are imagined to propagate freely through the sample keeping phase coherence. Any scattering is triggered by boundaries only, giving rise to quantum interference, which is manifested in the transport properties. For a conductor of irregular shape, or whose shape induces a chaotic dynamic in the limit of ray optics, the conductance shows fluctuations with respect to tuning parameters, such as the magnetic field[1], Fermi energy[2], and from sample to sample. These fluctuations are universal and depend only on symmetry.

However, in an experiment at finite temperatures, the actual conductance through a chaotic quantum dot is found to be non-universal. Thermal smearing and lattice vibrations, which induce inelastic scattering by phonons, among other sources of dephasing, destroy the phase coherence[1,3]. In this sense, the quantum predictions for the fluctuations of the conductance at zero temperature seem to be inaccessible or at least very difficult to attain at the microscopic scale. The underlying wave nature of the electrons offers the possibility to find undulatory alternatives in which the temperature does not play a significant role. In addition, in the Landauer-Büttiker formalism, the electronic transport is reduced to a scattering problem[4]; there the conductance in units of the conductance quantum ($G_0 = 2e^2/h \approx 1/13$ k$\Omega$) is just the transmission coefficient through the sample. This makes classical wave systems good models to verify the theoretical predictions for coherent quantum transport[5].

The first efforts to demonstrate the universal statistics of transmission through chaotic systems were done in microwave cavities[6,7]. However, the dissipation and the imperfect coupling that occur in these systems[8-10] makes verification of the predictions of the coherent transmission through quantum systems at zero temperature[9] difficult. Only partial and indirect verification has been performed by removing the imperfect coupling effects through a normalization procedure[11]. These difficulties can be handled in mechanical wave systems[12] since it has been found that the quality factor in these systems can be orders of magnitude higher[13] than those found in superconducting microwave cavities[14] thus not affecting phase coherence[15].

Here we report on the realization of a two-dimensional elastic cavity with the shape of a non-symmetric, half Sinai billiard, with two waveguides which allows us to measure coherent transmission, as illustrated in Fig. 1. The elastic cavity is constructed on an aluminum plate of thickness $h$ and is open through the different modes (or channels) supported by the two waveguides of width $W$. Flexural vibrations are excited and detected using non-contact electromagnetic-acoustic transducers[16]. Direct transmission between both waveguides, one in front to each other, is diminished by locating the semicircle of the Sinai billiard as an obstacle between them. To prevent wave reflections at the end of the waveguides, passive vibration isolation systems are used; these consist of wedges covered by foam pads.

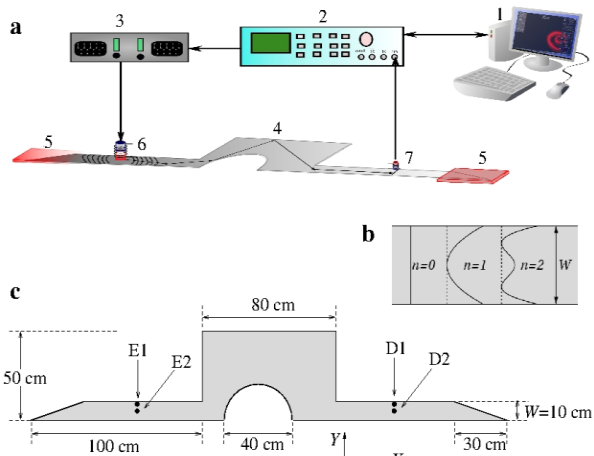

**Figure 1 | Experimental setup. a.** Workstation (1), vector network analyzer (2), high-fidelity audio amplifier (3), aluminum plate (4), passive vibration isolation systems (5), electromagnetic-acoustic transducers, exciter (6) and detector (7). **b.** Schematic drawing of the intensity of the lower bending channels in an elastic thin waveguide. **c.** Mechanical Sinai billiard on the aluminum plate of thickness $h$=6.35 mm (not shown); the waveguides are of width $W$=10 cm and the wedges at the end of them are covered by foams to avoid unwanted reflections. The points E1 and E2 (D1 and D2), indicate the used locations of the exciter (detector). Mechanical properties of the aluminum plate: the Young modulus, density, and Poisson's ratio are $E$=71.1 GPa, $\rho$=2,708 kg/m$^3$, and $v$=0.36, respectively.

The experimental setup consists of a workstation that controls a vector network analyzer (VNA) through a GPIB interface (see Fig. 1a). The VNA, Anritsu Model MS4630B, performs an up-chirp within the audible frequency range; a Cerwin-Vega high-fidelity audio amplifier CV-900 intensifies the signal coming from the VNA and sends it to an electromagnetic-acoustic transducer (EMAT). The EMAT, which consists of a coil and a permanent magnet, is located at one lead of the billiard. It excites flexural waves when the dipole moment axis of both the EMAT's coil and permanent magnet coincide with the normal to the plate[17]. The flexural waves travel along the elastic waveguide and enter into the thin-plate mechanical billiard, where they are scattered. A second EMAT of smaller dimensions, with the same configuration of coil and permanent magnet as the exciter, is located at the opposite waveguide, and is used to detect the flexural transmission. The signal measured by the EMAT detector is sent back to the VNA. The passive vibration isolation systems yield very good absorption for frequencies higher than 250 Hz, avoiding in this way undesirable stationary patterns. The low frequency motion for flexural waves in an elastic waveguide with free boundary conditions, is quantized into transverse modes as the wave propagates along the waveguide. The first three modes are sketched in Fig. 1b. In Fig. 1c the dimensions of the billiard and waveguides are depicted, where the points indicate the positions of the exciter (E) on one side of the cavity and the detector (D) on the opposite side. The four possible exciter-detector configurations are: E1-D1, E1-D2, E2-D1 and E2-D2, where E1 and D1 are 1 cm from the edge of the waveguide, and E2 and D2 are 5 cm from the edge. These positions are 40 cm away from the cavity.

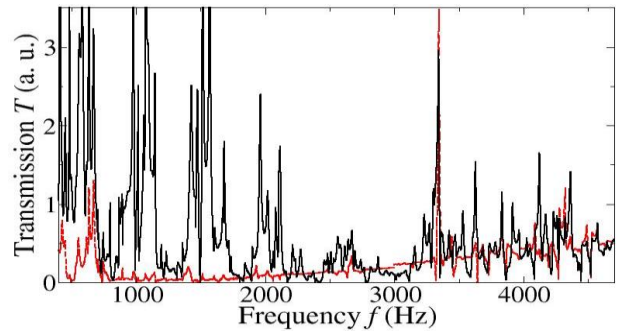

**Figure 2 | Transmission spectrum.** The transmission, in arbitrary units (a.u.), is measured at D1 (black line) and D2 (red line) when the excitation is at E2. Below 3,300 Hz only the mode $n$=1 is transmitted despite of both modes, $n$=0 and $n$=1, are excited. Above 3,300 Hz a different behavior is observed, which corresponds to the opening of the mode with $n$=2.

Figure 2 shows the measured transmission spectrum for two configurations, exciting at the center (E2) and detected at D1 and D2. One can notice that, for frequencies less than 3,300 Hz, the amplitude of the signal at the center of the waveguide (D2) is negligible with respect to that at the edge (D1). This means that although both modes $n$=0 and $n$=1 are open mainly the second one is transmitted as can be observed in Fig. 3, where the shapes of the measured modes are shown across the width of the waveguide. Evidence of the opening of the next mode (channel $n=2$) is observed around 3,300 Hz above which a new plateau appears. This is reminiscent of the *quantization of the flexural transmission,* being the flexural-wave equivalent of the quantization of the electrical conductance[18].

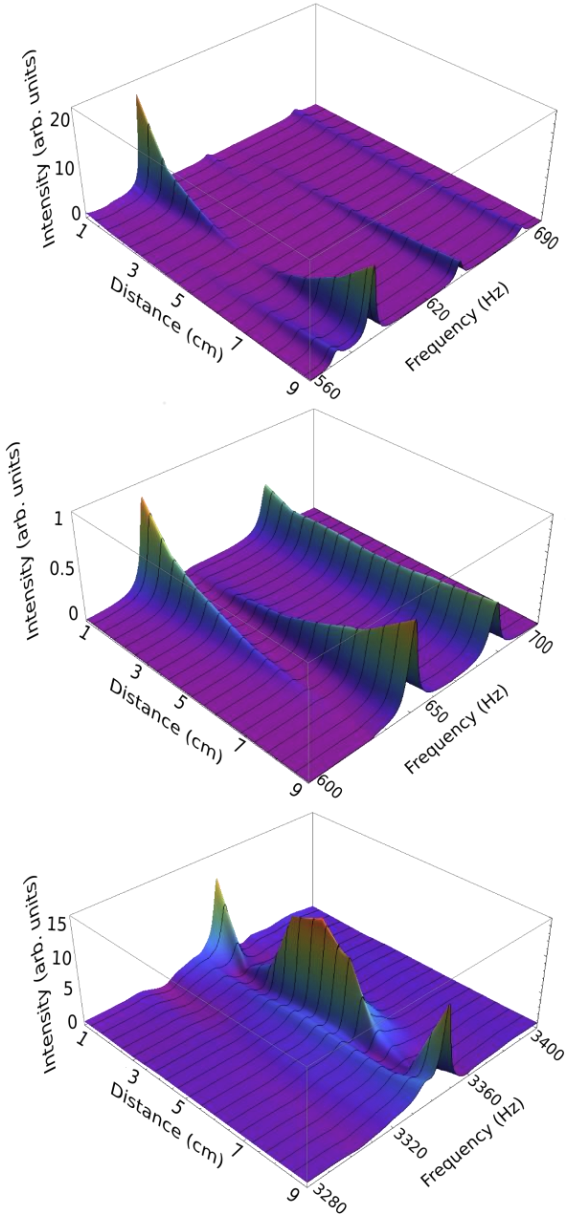

**Figure 3 | Measured modes.** Profiles of the transmitting bending modes. In the upper panel a mode with $n=1$ is observed at 580 Hz. The middle panel shows a mode with $n=0$ close to 676.8 Hz. Both modes are present below 3,300 Hz. A mode with $n=2$ is observed at 3,360 Hz (lower panel). At 640 Hz and 660 Hz a mix of modes with $n=0$ and $n=1$ appear; they show a very small amplitude. The transmission intensities were measured across the width of the waveguide.

Furthermore, over the smooth part of the transmission, a fluctuating part is observed., These fluctuations are analogous to those appearing in mesoscopic quantum dots[2] and in microwave cavities[6,8].

Since the mode with $n=1$ is mainly transmitted, the configurations E1-D1 and E1-D2, where the excitation is close to the edge, are the best suited for the statistical analysis of the transmission fluctuations. In Fig. 4 we show the transmission distributions for two frequency ranges. Assuming an ergodic hypothesis we compare the experimental results with the theoretical predictions from the random matrix theory. For frequencies between 110 Hz and 210 Hz, it is mainly the mode $n=1$ that is transmitted and the corresponding histogram of the transmission is in good agreement with the prediction of random matrix theory for the one channel case, $P(T)=1/2\sqrt{T}$. For frequencies in the interval 5,540-6,620 Hz, the modes with $n=1$ and $n=2$ are open and the experimental distribution also shows a good agreement with the theoretical predictions[5]. These results are very important since they verify for the first time the theoretical predictions of the simplest model for quantum transport by direct measurement of mode-to-mode transmission.

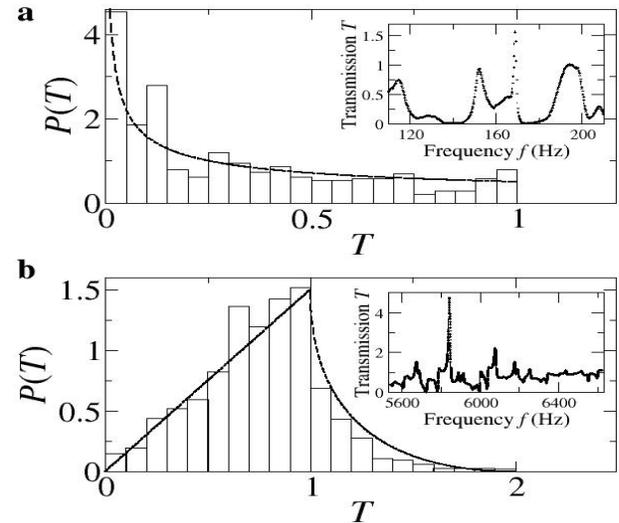

**Figure 4 | Transmission distributions.** Experimental distributions for the flexural transmission (histograms) show excellent agreement with the predictions of the simplest model of random matrix theory for quantum transport (dashed lines). **a.** One-channel case. **b**. Two-channels case. Insets: Transmission spectra normalized to the theoretical average in the corresponding frequency window. The histograms do not take into account the high and thin resonances appearing in the insets since they correspond to in-plane modes.

We have presented measurements of the coherent transmission of mechanical waves through a two-dimensional elastic chaotic cavity with two waveguides. We observed the quantization of the flexural transmission that occurred when each new mode opens. The statistics of the transmission agree with the theoretical predictions of the simplest model for quantum transport, something that had not been achieved before despite greats efforts in the subject. The transmission distribution for two open channels shows better agreement with the theory than the one-channel case since there are more resonances within the frequency window. Similar results were observed for different frequency intervals. This demonstrates the universal character of the transmission fluctuations for coherent transport. The results shown here open the possibility to study other quantum transport phenomena without decoherence.

**Acknowledgements** This work was supported by DGAPA-UNAM under project PAPIIT IN103115. E. Flores-Olmedo was supported by a scholarship from CONACYT. We acknowledge the kind hospitality of Centro Internacional de Ciencias A. C. and Benemerita Universidad Autonoma de Puebla for group meetings celebrated frequently there. We would like to thank A. Arreola, A. Fernández-Marín, V. Domínguez-Rocha and E. Sadurní for invaluable comments.